\documentclass[aps,prc,reprint,superscriptaddress]{revtex4-1} 

\usepackage{array} 
\usepackage{paralist} 
\usepackage{subfig} 
\usepackage{amssymb}
\usepackage{amsmath}
\usepackage{color}
\usepackage{graphicx} 
\usepackage{tabularx, ragged2e} 
\usepackage{epsfig}
\usepackage{tablefootnote}

\newcommand{\sci}[2]{\ensuremath{#1\!\times\!10^{#2}}}

\AtBeginDocument{\usepackage{booktabs}}

\begin{document}

\title{Detection of atomic nuclear reaction products via optical imaging}

\date{\today}

\author{B. Loseth}
\email{loseth@nscl.msu.edu}
\affiliation{National Superconducting Cyclotron Laboratory, Michigan State University, East Lansing, Michigan 48824, USA}
\author{R. Fang}
\affiliation{National Superconducting Cyclotron Laboratory, Michigan State University, East Lansing, Michigan 48824, USA}
\author{D. Frisbie}
\affiliation{National Superconducting Cyclotron Laboratory, Michigan State University, East Lansing, Michigan 48824, USA}
\author{K. Parzuchowski}
\affiliation{JILA, University of Colorado, Boulder, Colorado 80309, USA}
\author{C. Ugalde}
\affiliation{University of Illinois at Chicago, Chicago, Illinois 60607, USA}
\author{J. Wenzl}
\affiliation{Institute of Physics, Johannes Gutenberg University, D-55128 Mainz, Germany}
\author{J. T. Singh}
\affiliation{National Superconducting Cyclotron Laboratory, Michigan State University, East Lansing, Michigan 48824, USA}

\begin{abstract}
In this paper we propose a new method for measuring the cross section of low yield nuclear reactions by capturing the products in a cryogenically frozen noble gas solid. Once embedded in the noble gas solid, which is optically transparent, the product atoms can be selectively identified by laser induced fluorescence and individually counted via optical imaging to determine the cross section. Single atom sensitivity by optical imaging is feasible because the surrounding lattice of noble gas atoms facilitates a large wavelength shift between the excitation and emission spectrum of the product atoms. The tools and techniques from the fields of single molecule spectroscopy and superresolution imaging in combination with an electromagnetic recoil separator, for beam and isotopic differentiation, allow for a detection scheme with near unity efficiency, a high degree of selectivity, and single atom sensitivity. This technique could be used to determine a number of astrophysically important nuclear reaction rates.
\end{abstract}

\maketitle


\section{Motivation}

In stars and during stellar explosions, and over billions of years, intricate networks of nuclear reactions synthesized nearly every natural chemical element that we observe around us. Nucleosynthesis of most elements heavier than iron are not produced by stellar fusion but rather by neutron capture, whether it be slow and gradual individual neutron captures during stellar burning (s-process), or rapid capture of many neutrons such as is believed to occur during neutron star mergers (r-process). There are 35 stable isotopes inaccessible to neutron-capture processes and are believed to be produced through $\gamma$-induced photodisintegration (p-process) \cite{Burbidge1957, Meyer1994, Rauscher2013, Thielemann2017}.

There are a significant number of nuclear reactions that have a strong influence on nuclide abundances and whose cross sections are either unknown or poorly understood at astrophysically relevant energies. Measuring these cross sections is often technically challenging for a variety of reasons. At astrophysical energies (in the so-called Gamow window), the cross section can be extremely small due to the difficulty in overcoming the Coulomb barrier at stellar temperatures. In order to measure extremely small cross sections directly and within an acceptable time period, high beam currents and dense targets are required for the production of only a handful of reactions per day. 

An inverse kinematics configuration is often utilized, where the beam nuclei have higher mass than the target nuclei so that the reaction products scatter forward in a narrow cone. Based on their charge and mass, the few product nuclei are then separated from the beam and secondary nuclei by electric and magnetic fields in recoil separator systems \cite{Couder2008,Berg2018}. Alternative and often complementary methods involve detection of the proton, neutron, or gamma created by the reaction with an array of scintillating detectors around the reaction site \cite{Jaeger2001}. Unfortunately such methods are sensitive to cosmic ray, natural, and beam-induced background sources, the rates of which can be significantly higher than the reaction of interest. Some experimental efforts have moved deep underground, where the cosmic ray induced background rates are significantly lower. For underground facilities, CASPAR at Sanford Underground Research Facility \cite{Robertson2016}, and LUNA in Italy \cite{Formicola2014}, background rates become limited by radioactive elements in the surrounding rock, and are  $10^2 - 10^4$ times smaller than on the surface.

For reactions involving rare isotopes, it can be difficult to achieve sufficient statistics due to inadequate beam intensities. Rare isotope beams can also be significantly contaminated with other nuclei as a consequence of production mechanisms, which can drastically increase background rates. Furthermore, heavy nuclei have substantial magnetic rigidity and relatively slight differences in charge to mass ratios, making them cumbersome to separate due to the long distances and high magnetic fields required. Typical recoil separators are less effective at high masses for the same reasons.

\begin{figure*}[htp]
\centering
\includegraphics[width=\textwidth]{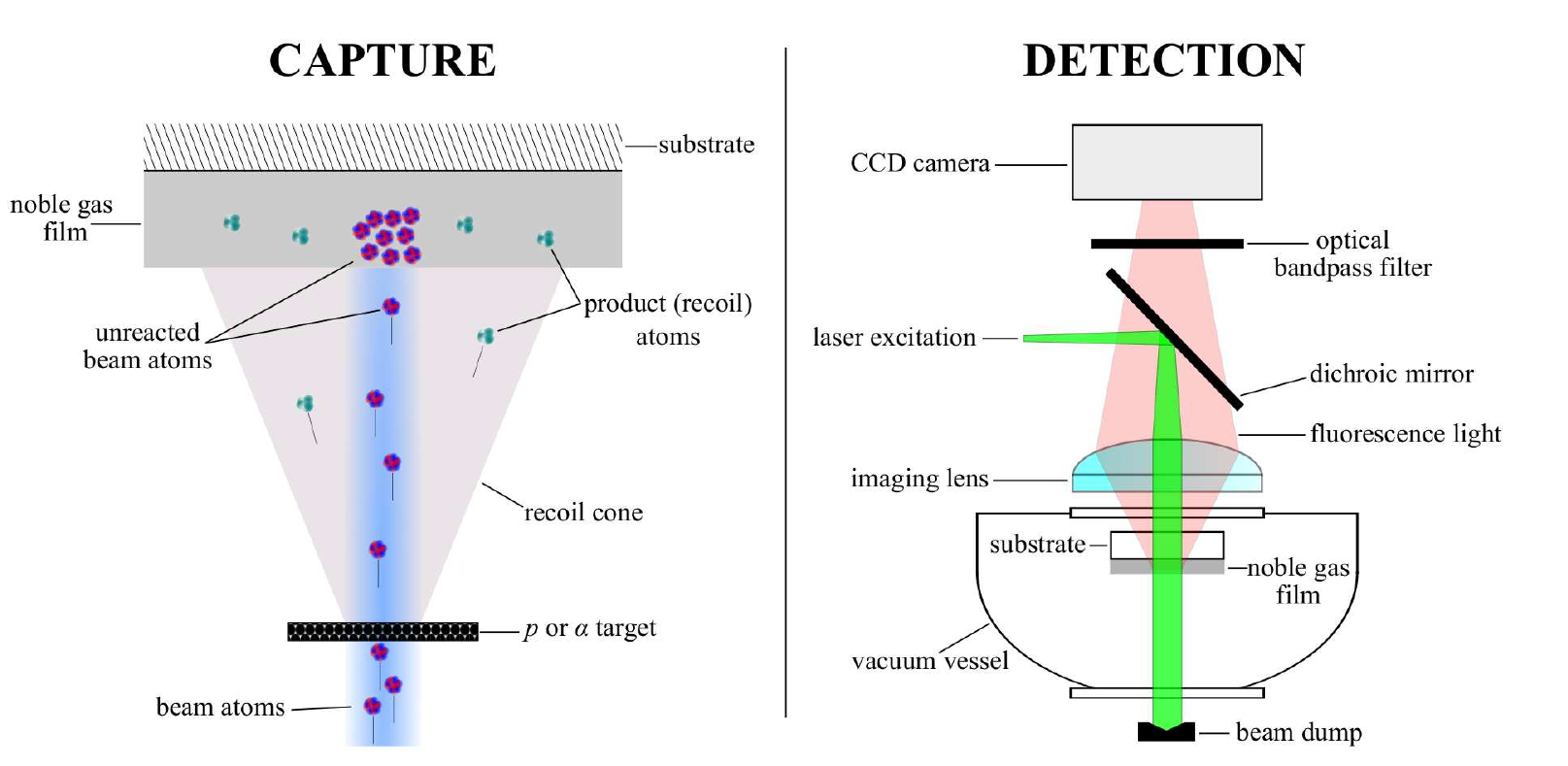}
\caption{ Graphical representation of the SAM concept (not to scale, noble gas film thickness exaggerated for clarity). left: Basic capturing scheme without a recoil separator. The nuclear reaction takes place in inverse kinematics, where the recoiling products and low intensity unreacted beam are captured in a noble gas film. right: Schematic of optical excitation and fluorescence imaging of the captured recoil atoms onto a CCD camera. The excitation light is separated from the emitted fluorescence light using optical bandpass filters.}
\label{fig:concept}
\end{figure*}

Novel detection schemes capable of bypassing the aforementioned challenges create an opportunity to measure exceptionally low yield nuclear reactions or other such low yield nuclear events, such as neutrinoless double-beta decay. Such a detection scheme should exhibit single atom sensitivity to the reaction products while being unsusceptible to traditional background sources. The detection methods should exhibit a high degree of selectivity between atomic species to overcome beam contamination or separation issues. A large detection efficiency is also highly desireable to maximize the probability of detecting rare events. 

We propose a technique for measuring cross sections of low yield nuclear reactions by detecting the atomic products \emph{optically}, called the Single Atom Microscope (SAM). The SAM is intended for reactions performed in inverse kinematics, such that most or all of the recoiling product atoms are captured inside a cryogenically frozen noble gas solid (such as Neon, Argon, etc.) deposited on a transparent substrate. Once trapped, an atomic resonance is excited in the product atoms with a laser, and the emitted fluorescence light is collected by a CCD camera-based imaging system. Guest atoms that are isolated in a noble gas matrix generally exhibit blue-shifted absorption and red-shifted emission bands relative to the wavelength in vacuum for a given atomic transition. This wavelength shift between absorption and emission bands (Stokes shift \cite{Rost1992}) can be as large as hundreds of nanometers. To detect the isolated product atoms, their redshifted fluorescence is isolated from transmitted or scattered excitation light with optical filters to pick out the emission wavelength range of interest.

A schematic of the method is included in Figure \ref{fig:concept}. We argue that single atom sensitivity is feasible with the SAM, noting that single atom detection of Barium atoms in solid Xenon has been demonstrated \cite{Chambers2018,Moe1991}. The main advantage of this approach is that the detection mechanism is not affected by traditional background sources. Neutron and gamma ray backgrounds do not affect the fluorescence spectra or detection thereof, and the product atoms are identified by their unique atomic transitions, which are distinct from any codeposited atoms from the beam. 

In the following section we discuss promising cross section sensitivities for two classes of nuclear reactions. We then go on to describe specific details regarding the capture and detection of atomic species in a noble gas solid. Finally, the method is summarized and the benefits and limitations of the SAM detection scheme are discussed.


\section{Cross Section Sensitivity}

\begin{table*}[htp]
\centering
  \caption{Candidate reactions for the Single Atom Microscope, with approximate beam currents, target areal densities, and expected yield.}\label{tab:reactions}
\begin{tabularx}{0.75\textwidth}{ l c c c c l }
\toprule
Reaction						& Beam		& Target		& Cross 		& Approx.		& Reaction				\\
							& current		& density		& section 		& yield		& Importance				\\
							& (pps)		& (atoms/cm$^2$)	& (barn)		& (products/day)	&						\\
\midrule
$^{22}$Ne$(\alpha,n) ^{25}$Mg 		& $10^{15}$	& $10^{19}$	& $10^{-15}$		& 1			& s-process $n$ source 			\\
					 		& $10^{15}$	& $10^{17}$	& $10^{-11}$		& 100			& 						\\
\addlinespace	
$^{91}$Nb$(p,\gamma)^{92}$Mo		& $10^4$		& $10^{20}$	& $10^{-5}$			& 1			& production of p-nuclei $^{92}$Mo	\\
							& $10^7$		& $10^{20}$	& $10^{-3}$		& $10^5$		& 						\\
\addlinespace
\bottomrule
\end{tabularx}
\end{table*}

We envision two classes of reactions where the SAM detection scheme is applicable: 1) extremely small cross section reactions with a high current stable isotope beam, and 2) low current rare isotope (radioactive) beam reactions. Table \ref{tab:reactions} contains approximate experimental parameters for example reactions of each type.

{\bf Small cross section reactions.} $^{22}$Ne$(\alpha,n)^{25}$Mg is a key reaction for s-process nucleosynthesis, and it has an extremely small predicted cross section in the Gamow window on the order of femtobarns ($10^{-15}$ barn) \cite{Wiescher2012}. Assuming a high intensity $^{22}$Ne beam of current $10^{15}$ pps incident on a windowless $^4$He gas jet target with an areal density of $10^{19} $ atoms/cm$^{-2}$ (JENSA target \cite{Chipps2014}), the expected yield for a 1 fb cross section is only a single $^{25}$Mg atom per day. Due to such small yields, single atom sensitivity to the product $^{25}$Mg atoms with negligible background rates are necessary to measure a cross section for this reaction in a reasonable amount of time, even at the highest achievable currents and target densities. 

For the SAM to measure this reaction, it should be noted that such high beam currents require that the unreacted beam intensity ($10^{15}$ pps) be attenuated by a factor of $10^6$ in order to avoid melting a cryogenic noble gas film, as will be discussed in in Section III. With an appropriate recoil separator to attenuate the unreacted beam intensity, this reaction is well suited for the SAM as the beam ($^{22}$Ne) is a noble gas and thus any unseparated $^{22}$Ne beam atoms would not contribute background fluorescence during optical imaging of the product $^{25}$Mg atoms. An advantage is that, unlike some traditional detection methods, the SAM detection scheme would be immune to leaky beam or other non-magnesium beam contaminants that can be difficult to completely eliminate from high intensity beamline systems. It is furthermore desirable to have an alternative technique, that the SAM detection scheme can potentially satisfy, with different systematics than the forthcoming underground measurements due to the importance of this reaction.

{\bf Low beam current reactions.} The low intensity of rare isotope beams is often the limiting factor in the cross section sensitivity for traditional detection methods. With single atom sensitivity and atomic species selectivity, the SAM scheme can improve upon cross section sensitivities even with low intensity beams. As an example, the reaction $^{91}$Nb$(p,\gamma)^{92}$Mo has been identified as a key reaction in the production of the p-process nuclei $^{92}$Mo \cite{Rauscher2016}. The NON-SMOKER database reports cross sections from 1 $\mu$b to 1 mb in the Gamow window \cite{Rauscher}. At the National Superconducting Cyclotron Laboratory (NSCL), a $^{91}$Nb beam can be produced with a current of order $10^4$ pps, which will be further improved to $10^7$ pps at the upcoming Facility for Rare Isotope Beams (FRIB). An advantage of such small currents for SAM is that no beam rejection or recoil separator would be necessary to protect the noble gas film, easily allowing for capture efficiency approaching unity. Using a proton target (CH$_2$, of density 0.25 mg/cm$^2$), for a 10 $\mu$b cross section roughly one $^{92}$Mo atom is produced daily with NSCL beam intensities, and the same yield allows for 10 pb cross sections to be probed with FRIB beam currents assuming negligible background rates.

Gamma-based detection methods are handicapped by lower efficiencies and are susceptible to cosmic ray and environmental gamma sources, requiring more beam time to achieve sufficient counts above background (typically hundreds of counts for good statistics). Similar difficulties exist for low energy neutron detection. Reactions involving higher mass nuclides, such as $^{91}$Nb$(p,\gamma)^{92}$Mo, easily exceed the design magnetic rigidity acceptance of contemporary recoil separator systems \cite{Couder2008,Berg2018} unless a high charge state is selected, a limitation on the overall efficiency despite near unity detection efficiency of recoils after the separator. The SAM has the potential to significantly outperform gamma, neutron, and electromagnetic separator-based methods for some rare isotope reactions due to smaller beam time requirements to amass sufficient statistics, especially at lower cross sections where expected yields are small.


\section{Technical Challenges}

A number of technical challenges must be overcome before this method can be applied to measuring the cross section for a low yield reaction. Chiefly among them, single atom sensitivity must be demonstrated for the product species of interest, which we argue is feasible for many species. Achieving single atom sensitivity requires performing time-dependent calibrated fluorescence spectroscopy of the species of interest in a solid noble gas, as well as optical background characterization at excitation and emission wavelengths appropriate to the product species to be detected. As laser intensity and optical requirements may not allow for imaging the entire area containing product atoms simultaneously, a laser scanning system will be implemented to raster across the surface of the deposited solid noble gas film. Furthermore, single atom detection should be achieved for short optical integration times, so the optical signal to background rate should be maximized to ensure that imaging the entire substrate via rasterized scanning is not prohibitively time-consuming. There are also a few outstanding questions regarding capture and neutralization of energetic ions in a cryogenic solid noble gas.


\subsection{Capture in noble gas solids}

To capture the products, generally speaking, a solid noble gas film of thickness 100 $\mu$m  is sufficient to fully stop an ion with a kinetic energy of a few MeV/nucleon \cite{Ziegler2010a}, which is at the higher end of the energy range for most reactions of astrophysical interest. Highly transparent thin films of thickness of 100 $\mu$m can be deposited in around an hour with an area 20 cm$^2$ or larger, which matches the size of the focal plane for a typical recoil separator. The specific properties of most noble gases in solid form are contained in Table \ref{tab:ngs}. The selection of which noble gas to use for a given reaction will depend primarily on the matrix-isolated spectra of the product atoms to be detected. The polarizability of the noble gas atoms has a significant effect on the spectra of trapped atoms \cite{Gerhardt2012}.

There are a few important factors to consider regarding the capture of energetic ions in a noble gas solid. First, some amount of damage will be inflicted on the noble gas film through direct heating and surface sputtering due to exposure to an energetic ion beam. Second, all product atoms are highly ionized and may not be completely neutralized before stopping in the film. As ions may have drastically different spectra than neutral atoms, the fraction of product atoms which remain ionized may be optically undetectable. Third, it is unclear what trapping site the stopped atoms will occupy in the face-centered cubic (fcc) lattice formed by noble gas atoms (called the noble gas matrix), and trapping sites are known to affect the spectra of the captured atoms \cite{crepin-gilbert1999}.

{\bf Noble gas film damage.} Two obvious mechanisms affect the maximum beam intensity at which significant damage is inflicted on the noble gas matrix. The kinetic energy of any unseparated beam atoms is deposited as heat in the noble gas film, which will cause the film to sublime for beam sufficient intensities. This effect may be especially significant for Neon due to the single digit temperatures required to solidification, where the cooling power of contemporary pulse-tube cryocoolers is only on the order of 1 Watt. The heavier noble gas films will be more resistant to direct-heating sublimation, as the cooling power improves to tens of Watts at higher temperatures. It is important to note that noble gas ices are electrical insulators, and will therefore have poor thermal conductivity at low temperatures, which may prove to be a limiting factor despite sufficient cooling power \cite{Ekin2006}. To get a sense of a typical heat load, a 3 MeV/nucleon $^{91}$Nb beam with a current of $10^{8}$ pps would deposit a tolerable 4 mW.

\begin{table}[tp]
\centering
  \caption{Properties of noble gas solids.}\label{tab:ngs}
\begin{tabularx}{0.9\columnwidth}{ l c c c c c }
\toprule
				& Ne 		& Ar 		& Kr 		& Xe 		& ref				\\ \midrule
lattice structure	 	& fcc		& fcc		& fcc		& fcc 		& \cite{JohnR.Rumble2018}	\\
lattice constant (\AA)	& 4.464	& 5.311	& 5.646	& 6.132 	& \cite{klein1977rare}		\\
triple point (K)		& 24.56	& 83.81	& 115.78	& 161.37 	& \cite{klein1977rare}		\\
$\rho_{solid}$, t.p. (g/cm$^3$) 		& 1.444	& 1.623	& 2.826	& 3.399 	& \cite{Pollack1964}		\\
T$_{solid}$ (K, \@ $10^{-6}$ Pa)	& 7.3		& 27.4	& 38.4	& 51.3 	& \cite{Shakeel2018}		\\
Sublim. energy (meV)	& 19.6	& 80		& 116		& 164		& \cite{Schou1987}			\\
polarizability (\AA$^3$)	& 0.394	& 1.641	& 2.484	& 4.044 	& \cite{JohnR.Rumble2018}	\\
refractive index  		& 1.11	& 1.29	& 1.38	& 1.49 	& \cite{Schulze1974}		\\

\bottomrule
\end{tabularx}
\end{table}

The likely more significant damage mechanism is surface sputtering of the film by the beam and products. Noble gas matrices are relatively loosely bound, and each incoming energetic ion will eject some number of noble gas atoms from the matrix, typically called the sputtering yield. This effect can be compounded under high beam intensities, as the ensuing higher temperatures due to kinetic energy deposition increases noble gas atom mobility and effectively lowers the surface binding energy. For light ions ($p$, $\alpha$) with kinetic energy on the order of MeV incident on sufficiently thick, low temperature noble gas films, the sputtering yield is determined by the sublimation energy and the electronic stopping power of the noble gas solid \cite{Brown1986,Schou1987}. 

The literature only reports sputtering of noble gas films by heavy ions at a low kinetic energy, in the range of 1-10 keV, where the sputtering yield is dominated by nuclear stopping power among other effects \cite{Schou1992,Balaji1990a}, in contrast to the light ion case. Balaji \emph{et al.} report sputtering yields as high as $10^3 - 10^5$  with 5 keV ions for various combinations of Ne, Ar, Kr, and Xe ions and targets \cite{Balaji1990a}. To get a more macroscopic understanding of this effect, under beam intensity of $10^9$ ions/cm$^2$/s, such sputtering yields correspond to a thickness loss of roughly 0.002 - 0.2 $\mu$m/hr. Studies have not been performed for medium to high mass ions impingent on noble gas solids at astrophysical energies of a few MeV/nucleon, where electronic stopping power will be dominant, and where electronic stopping powers are an order of magnitude higher than the light ion case. Extrapolating the low energy heavy ion sputtering yields to astrophysical energies, maximum thickness loss would increase to 2 $\mu$m/hr, assuming sputtering yields are proportional to total stopping power.

{\bf Product ion neutralization and trapping.} For studies of the optical spectra of atomic species in noble gas matrices, samples are typically prepared by depositing an initial layer of the noble gas matrix on the cooled substrate, followed by a layer of co-deposited noble gas and guest species, and finished with a final layer of noble gas to ensure that each guest atom is isolated (i.e. surrounded by noble gas atoms). The guest species are usually deposited with either an effusive or ionic source; studies of Na$^{+}$ ions deposited in Ar \cite{Silverman1997} and Ba$^{+}$ ions in Xe \cite{Mong2015} have shown the spectra consistent with their neutral counterparts, and it is known that the charge state of energetic ions stopped in medium approaches zero \cite{Betz1972}. However, as noble gas solids have poor electrical conductivity, it is unknown whether there is a sufficient population of loosely bound electrons for complete neutralization. Furthermore, it may be advantageous for some species to remain singly ionized due to more favorable spectroscopy.

\begin{table*}[htp]
\footnotesize
\centering
  \caption{Selected Matrix-Isolated Absorption and Emission Spectra of SAM-Friendly Species.}
  \label{tab:atoms}
\begin{tabularx}{0.77\textwidth}{ c c c l l l l c }
\toprule

	 & 	& \multicolumn{2}{c}{vacuum transition}				& \multicolumn{2}{c}{matrix isolated}				&			& 	\\
\cmidrule(lr){3-4}\cmidrule(lr){5-6}

Atom 	& Z $\quad$ 	& assignment 			& $\lambda$ (nm) $\quad\quad$	& absorption $\lambda$ (nm) \footnote[1] 	& emission $\lambda$ (nm) 	& lifetime (ns)\footnote[2]	& ref	\\
\midrule
		
Li	& 3 	& 2p $^2$P$\leftarrow$2s $^2$S 	& 671.0 			& Ar 656.5-679.0\footnote[3]		& 900 				& \emph{26}		& \cite{Wright1980c}			\\
\addlinespace
Na	& 11 	& 3p $^2$P$\leftarrow$3s $^2$S 	& 589.2, 589.8		& Ar 536.0-594.5\footnote[3]		& 670-710			& $13-28$		& \cite{Balling1978a}				\\
\addlinespace
K	& 19 	& 4p $^2$P$\leftarrow$4s $^2$S 	& 766.7, 770.1		& Ar 666.4-746.7\footnote[3]		& 850-950			& $20-75$		& \cite{Balling1979a}				\\
\addlinespace
Rb	& 37 	& 5p $^2$P$\leftarrow$5s $^2$S 	& 780.2, 795.0		& Ar 705-755\footnote[3]			& 877				& $\sim20$		& \cite{Gerhardt2012,Balling1983a}	\\
	&  	& 4d $^2$D$\leftarrow$5s $^2$S 	& 516.5			& Ar 420-540			& 630				& $\sim100$		& \cite{Balling1983a}				\\
	&  	& 6p $^2$P$\leftarrow$5s $^2$S 	& 420.5			& Ar 420-540			& 630				& $\sim100$		& 						\\
\addlinespace
Cs	& 55 	& 6p $^2$P$\leftarrow$6s $^2$S 	& 852.3, 894.6		& Ar 822-845\footnote[3]			& 970				& \emph{30.5}, \emph{35}	& \cite{Balling1983a}		\\
	&  	& 5d $^2$D$\leftarrow$6s $^2$S 	& 685.1			& Ar 610-670\footnote[3]			& 762				& $\mathit{2.2\times 10^{10}}$	& 				\\
	&  	& 7p $^2$P$\leftarrow$6s $^2$S 	& 455.6			& Ar 440-520\footnote[3]			& 762				& \emph{543}			&				\\
\addlinespace
Be	& 4 	& 2p $^1$P$\leftarrow$2s $^1$S 	& 234.9 			& Ne 232.0				& 232 				& \emph{1.8}	& \cite{Brom1975}				\\
	&  	&  						&  				& Ar 235.0-237.0\footnote[3] 		& 465				& \sci{1.33}{9}	& 						\\ 
	&  	&  						&  				& Kr 240.5				& 464.7			& \sci{9.5}{7}	& 						\\
\addlinespace
Mg	& 12 	& 3p $^1$P$\leftarrow$3s $^1$S 	& 285.3 			& Ne 275.3\footnote[3]			& 296.5			& 2.03		& \cite{Healy2012a}			\\
	&  	&  						&  				& Kr 277.0-296.0\footnote[3] 		& 297-326			& $1.25-2.25$	& \cite{McCaffrey1994a}			\\ 
	&  	& 3p $^3$P$\leftarrow$3s $^1$S	&  				& 			 		& 472				& \sci{8.91}{6}	& 						\\ 
\addlinespace
Ca	& 20 	& 4p $^1$P$\leftarrow$4s $^1$S 	& 422.8 			& Ar 422.0				& 432.9			& \emph{4.6}	& \cite{Bondybey1978}			\\
	&  	& 4p $^3$P$\leftarrow$4s $^1$S	& 	 			& 					& 647.6			& \sci{8.6}{5}	& 						\\
\addlinespace
Sr	& 38 	& 5p $^1$P$\leftarrow$5s $^1$S 	& 460.9			& Ar 447				& 466.2			& \emph{5}		& \cite{Miller1978}				\\
	&  	& 5p $^3$P$\leftarrow$5s $^1$S	& 689.4 			& 					& 709.2			& $\mathit{2.1\times 10^{4}}$	&				\\
\addlinespace
Ba	& 56 	& 6p $^1$P$\leftarrow$6s $^1$S 	& 553.7 			& Ar 532				& 550				& \emph{8.4}	& \cite{Mong2015}				\\
	&  	& 					 	& 	 			& Xe 561-566			& 570-591			& 			& 						\\
\addlinespace
Zn	& 30 	& 4p $^1$P$\leftarrow$4s $^1$S 	& 213.9 			& Ne 205.4				& 212.8			& 1.15		& \cite{Healy2012a}			\\
	&  	&  						& 	 			& Xe 219.9				& 356, 399			& $>10^4$		& \cite{Bracken1997}			\\
	&  	& 4p $^3$P$\leftarrow$4s $^1$S 	& 307.7 			& Ar 297				& 				& $\mathit{2.6\times 10^{4}}$	& \cite{Laursen1991}	\\
\addlinespace
Cd	& 48 	& 5p $^1$P$\leftarrow$5s $^1$S 	& 228.9 			& Ne 216.5-221.7$^a$		& 227.2			& 1.26		& \cite{Healy2012a}			\\
	&  	& 5p $^3$P$\leftarrow$5s $^1$S 	& 326.2 			& Ar 312.4				& 				& $\mathit{2.5\times 10^{3}}$	& \cite{Laursen1991}	\\
\addlinespace
Hg	& 80 	& 6p $^3$P$\leftarrow$6s $^1$S 	& 253.7 			&  Xe 253.2				& 				& \emph{119}	& \cite{Laursen1991}			\\
\addlinespace
Al	& 13 	& 3d $^2$D$\leftarrow$3p $^2$P 	& 308.3 			& Ne 260.0 				&  				& \emph{17}		& \cite{Ammeter1973,Grinter1987}	\\
	&  	& 4s $^2$S$\leftarrow$3p $^2$P	 	& 394.5 			& Ne 320.0				& 				& \emph{20}		& 						\\
\addlinespace
S	& 16 	& 3p $^1$S$_0\leftarrow$ 3p $^3 $P$_1$	& 459.0			& Ar 456.9				&  				& $\mathit{3.3\times 10^9}$	& \cite{Khriachtchev1998}	\\
	&  	& 3p $^1$S$_0\leftarrow$ 3p $^1 $D$_2$	& 772.7			&  					& 785				& \sci{2.3}{5}	& 						\\
\addlinespace
Mo	& 42 	& 5p z $^7$P$\leftarrow$ 5s a $^7$S	& 379.8			& Ar 341.3				& 399.0			& $>10^3$		& \cite{Pellin1983}				\\
	&  	& 5s b $^5$D$\leftarrow$ 5s a $^7$S	& 				& 			 		& 496.8			& \sci{1.5}{5}	&						\\
\addlinespace
Yb	& 70 	& 6p $^1$P$\leftarrow $ 6s $^1$S	& 398.8			& Ne 388.2 				& 394.9			& \emph{5.2}	& \cite{Xu2015}				\\
	&  	& 6p $^1$P$\leftarrow $ 6s $^1$S	& 555.8			& 	 				& 546.0			& \sci{6.8}{2}	& \cite{Xu2014}				\\
\bottomrule
\end{tabularx}
\vspace*{-2.5cm}
\footnotetext[1]{linewidths in medium are typically on the order of $1-10$ nm}
\footnotetext[2]{\emph{italicized values} are vacuum lifetimes, NIST Atomic Spectra Database (physics.nist.gov)}
\footnotetext[3]{multiplet pattern observed}
\end{table*}

The implantation mechanism for energetic nuclear reaction product ions, which will penetrate some depth into the film, is starkly different than the typical preparation method. In particular, the ions will be highly ionized and it is not clear what the precise ratio of incoming ions will become fully neutralized before stopping in the noble gas film. Furthermore, the trapping site of the stopped product atom in the noble gas atom lattice may be unstable or significantly different than the trapping sites for typical noble gas matrix samples. Fortunately, annealing noble gas films has been shown to recover atoms in unstable trapping sites \cite{crepin-gilbert1999}. These questions require further investigation as they directly limit the SAM detection efficiency.


\subsection{Optical signal to background estimates}

After capture, the product atoms must be identified and detected in the noble gas film based on their atomic spectra. It is advantageous that the spectral behavior of atoms and molecules isolated in noble gas matrices has been a field of study in chemical physics for decades, and so the spectra of many atomic species have already been measured in a variety of matrices. Broadly speaking, the transitions of atomic species isolated in noble gas matrices are qualitatively similar to transitions in vacuum, however transition wavelengths can be shifted by tens to hundreds of nanometers (nm) and exhibit significantly broadened linewidths (typically $1-10$ nm). Table \ref{tab:atoms} contains a subset of the available atomic spectra in noble gas matrices along with vacuum transition wavelengths. The lifetimes of allowed transitions are not significantly affected in medium \cite{Xu2014}, and so transitions lacking any available lifetime data in medium are listed with their vacuum lifetimes. This table is not exhaustive, as many species and transition data have been omitted for brevity, but it does include species compatible with the SAM detection scheme. 

The physics of atoms and their electronic spectra interacting with noble gas atoms is thoroughly reviewed in \cite{crepin-gilbert1999}. Our proposed optical detection scheme relies on the shift between excitation and fluorescence spectra exhibited by most species in medium (see Figure \ref{fig:concept}), which allows for the selective optical filtration of any transmitted or scattered excitation light. Divalent atoms in particular can exhibit considerable shifts due to an intersystem crossing behavior, such as Yb in Ne \cite{Xu2011}, Mg in Kr \cite{McCaffrey1994a}, and Hg in Ar and Kr \cite{Duley1967}, where the perturbative effect of the noble gas lattice facilitates a radiationless transition from an excited state to an adjacent or lower lying state. 

{\bf Single atom signal rate.} The net optical signal rate due to a single resonantly emitting atom is simply the fluorescence intensity $F$ (number of photons isotropically emitted per unit time) per atom multiplied by the efficiency of the optical imaging system, which we estimate to be on the order of $10^{-2}-10^{-3}$. Optical imaging efficiency contains factors due to the solid angle, transmission efficiency of optical filters for wavelength separation of the excitation from emission light (Semrock, Rochester, NY), and the wavelength-dependent quantum efficiency of CCD cameras. Laser coolable atoms are ideal, as they are generally characterized by having cycling transitions with no or minimial repumping. For an alkali atom under resonant excitation from the ground state $a$ to first excited state $b$, the fluorescence intensity $F$ is half the inverse of the excited state lifetime, assuming the excitation light is of sufficient intensity. Consulting Table \ref{tab:atoms}, the $^2S\rightarrow$$^2P$ transition of Rb atoms in solid Ar exhibit a 20 ns lifetime, corresponding to an optical signal rate of roughly $25-250$ kHz depending on imaging efficiency.

\begin{figure}[btp]
\centering
\epsfig{file=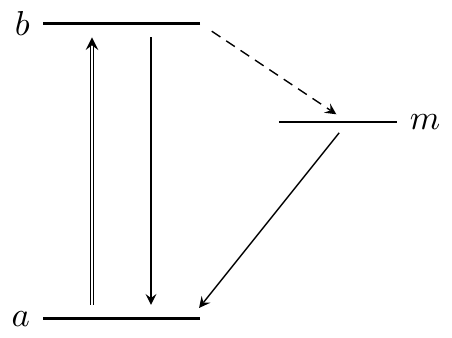,width=0.8\columnwidth}
\caption{Generic energy level diagram for a three level system with ground state $a$, excited state $b$, and metastable state $m$.  Excitation is labeled by double arrow ($\Rightarrow$), emission by single arrows ($\rightarrow$), and nonradiative transfer by dashed arrow ($\dashrightarrow$).}
\label{fig:eld}
\end{figure}

The alkaline earth elements, with two valence $s$ shell electrons, are slightly more complicated. As depicted in Figure \ref{fig:eld}, upon resonant excitation from $a \rightarrow b$, there is some chance to transfer from $b$ to a lower energy metastable state $m$ with significantly longer lifetime. Mg atoms in solid Kr exhibit a 2 ns lifetime for the $^1S \rightarrow$$^1P$ transition, corresponding to an optical signal rate of $0.25-2.5$ MHz. However, emission from the triplet $^3 P$ state was also observed with a 9 ms lifetime ($0.5$ Hz optical signal rate) \cite{McCaffrey1994a}. Detection of a Mg atom via the $^1 P$ emission appears feasible based on these lifetimes, as a sufficient number of photons will be detected before the atom transfers to the metastable $^3 P$ state. Waiting for the metastable state decay or repumping the atom with a secondary light source should allow for recovery of the $^1 P$ emission band. This blinking into and out of metastable states is characteristic of a single emitter, and observation of blinking behavior would go towards confirmation of single atom sensitivity. It should be noted that detection of the $^3 P$ emission is technically feasible, as the optical signal rates are still well above the dark count rate of order 1 mHz for state-of-the-art CCD cameras (Andor, Belfast, United Kingdom). Furthermore, background rates may be significantly lower at the 472 nm $^3 P$ Mg emission compared to the $297-326$ nm $^1 P$ emission, whether it be due to the effectiveness of optical filters for intense ultraviolet excitation light, the relative wavelength shift between excitation and emission bands, or fluorescence of impurities in the windows and optics.

Detection of a transition metal like Molybdenum is expected to be more challenging than the previous cases, with a $4d^5 5s^1$ electronic configuration and $a$ $^7S$ ground state. Studies of matrix isolated Mo in solid Ar and Kr by Pellin \emph{et al.} \cite{Pellin1983} report substantial nonradiative transfer to metastable states widely separated in gaseous Mo atoms despite spin, parity, or $J$ selection rules. Emission from metastable $b$ $^5D$, $a$ $^5P$, and $a$ $^5F$ states were observed with similar lifetimes after excitation to $z$ $^7P$ in an Argon matrix. Taking the reported in-medium lifetimes at face-value (Table \ref{tab:atoms}), observation of the $z$ $^7P$ fluorescence will yield kHz signal rates, while the metastable $b$ $^5D$ state would yield 30 Hz signal rates. The challenge becomes determining an excitation scheme that effectively mimics the three level system depicted in Figure \ref{fig:eld}, analogous to the Magnesium case.

\begin{table*}[htp]
\centering
  \caption{Potential sources of optical background, with known excitation wavelengths.}
\label{tab:backgrounds}
\begin{tabularx}{0.72\textwidth}{ c c c c }
\toprule
Background		& Location/Source	& Wavelength 			& Notes 				\\ \midrule
scattered light	& laser excitation 	& species dependent		& attenuate with optical filter(s)	\\
unreacted beam	& noble gas film 	& species dependent		& 	\\
beam contaminant	& noble gas film 	& species dependent		& 	\\
N$_2$			& film/residual gas 	& $<200$ nm 	 		&     off resonance			\\
N			& film/residual gas 	& 523, 1047 nm \cite{Niino2001}	&     unknown concentration	\\
O$_2$		& film/residual gas 	& 763 nm \cite{Becker1988}	& 1 nm FWHM. 24 ms lifetime     \\
O			& film/residual gas 	& 296, 558, 630 nm			&     unknown concentration	\\
H$_2$O		& film/residual gas 	& $<200$ nm 			&     off resonance			\\
C			& film/residual gas 	& 462, 872, 980 nm			& unknown concentration		\\
Cr$^{3+}$		& sapphire impurity & 693.0, 694.4nm \cite{He1997}	& impurity in substrate		\\
\bottomrule
\end{tabularx}
\end{table*}

{\bf Signal to background estimation.} Estimating the optical background rate is a more challenging task. The high number of possible optical background sources hampers the declaration of a general quantitative assertion about the background rate, and ultimately it will have to be measured and minimized for a given species through adjustments in optical spectroscopy geometry and materials selection. Instead, we will catalog some possible sources of background light and estimate their relative importance (see Table \ref{tab:backgrounds}). Any scattered or reflected excitation laser light is expected to be sufficiently attenuated through the use of optical filters. The primary sources of background light are expected to be contaminant atoms or molecules that, under the excitation wavelength of the product atom of interest, happen to fluoresce at wavelengths within the bandpass of the optical filters. These contaminants could be impurities in the substrate, noble gas film, vacuum windows, or optics. Furthermore, the beam can be contaminated by isotopes with similar charge to mass ratios that will be implanted alongside the product atoms.

The overall background rate will be related to the sum of the fluorescence rates for all background sources. Assuming the experimental conditions with excitation light at $500$ nm wavelength, with an intensity (power per unit area) of $P_\gamma / A_\gamma = 1 \text{W/cm}^2$, the optical signal to background ratio $S/B$ for one product atom with a $\tau=5$ ns excited state lifetime is approximately
\begin{align}
S/B = \left[\frac{2 \tau P_\gamma}{h \nu_\gamma} \sum_i n_i \sigma_0^i \left( \frac{\sigma^i(\nu_\gamma)}{\sigma_0^i} \right) \right]^{-1}& \\
 \approx 4 \times 10^{-14} \left[\sum_i n_i \sigma_0^i \left( \frac{\sigma^i(\nu_\gamma)}{\sigma_0^i} \right) \right]^{-1}&,
\end{align}
where the sum is over all background species with areal density $n_i$, peak absorption cross section $\sigma_0^i$, wavelength dependent absorption cross section $\sigma^i (\nu_\gamma)$  at a laser frequency $\nu_\gamma$, and $h$ is the Planck constant.

\begin{figure}[tp]
\centering
\epsfig{file=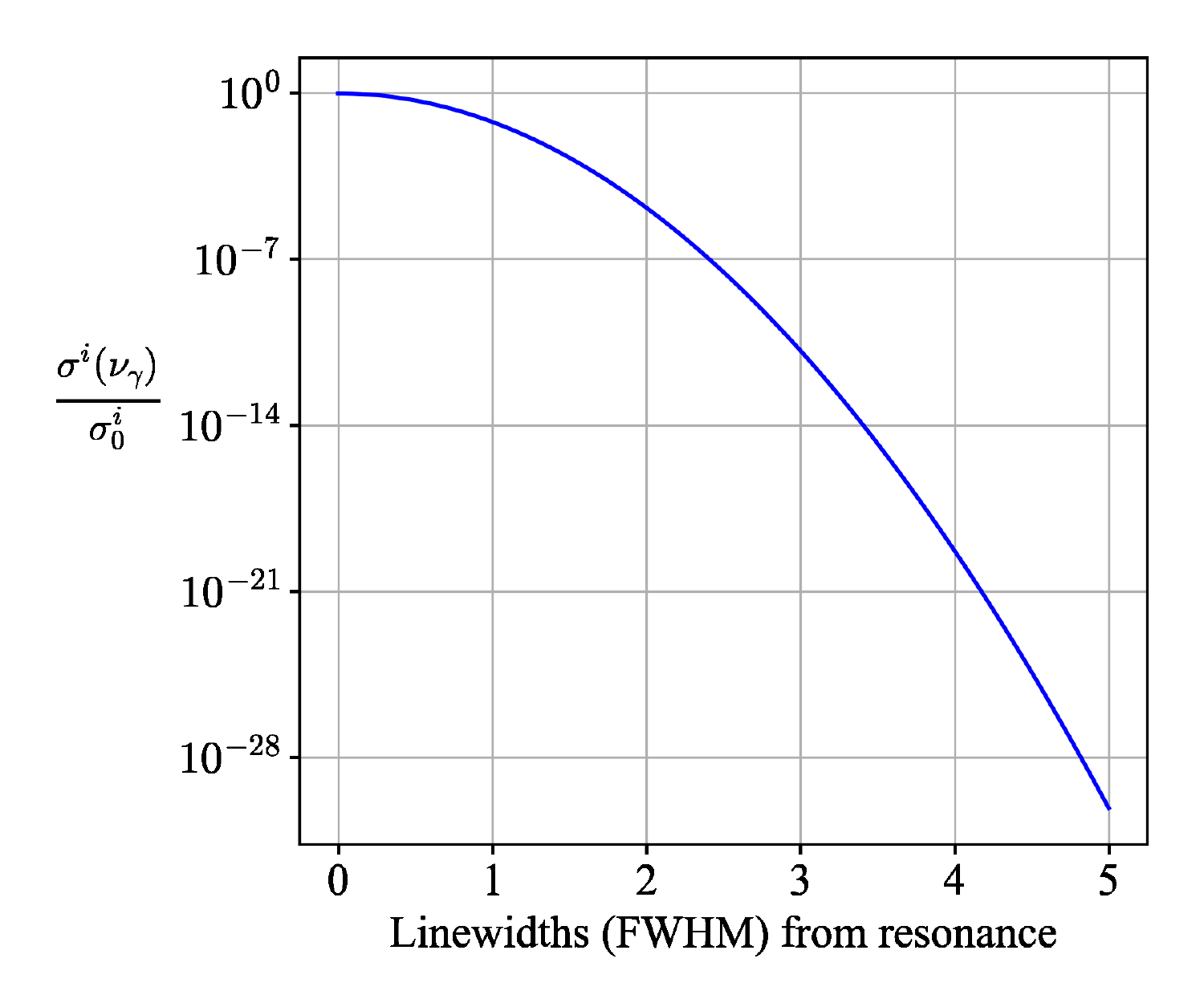,width=\columnwidth}
\caption{Off-resonance suppression factor $\sigma^i(\nu_\gamma)/\sigma_0^i$ for Gaussian absorption lineshapes. Far from resonance, the probability of excitation decreases exponentially, suppressing the probability of impurity fluorescence.}
\label{fig:offres-suppr}
\end{figure}

Equations (1-2) state that $S/B$ is inversely proprotional to the areal number density $n_i$ and the absorption cross section $\sigma^i (\nu_\gamma)$ of background species. The absorption cross section is dependent on the in-medium lineshape factor, which is a function describing the probability of absorption as a function of wavelength, and is typically Gaussian for matrix isolated species. Figure \ref{fig:offres-suppr} illustrates the off resonance suppression factor as a function of linewidths from the transition resonance, assuming the in-medium lineshape remains Gaussian far from resonance. If the peak absorption wavelength of a background species is sufficiently far from the excitation wavelength, the background atom excitation rate will be exponentially suppressed.  It is important to note that linewidths for absorption and emission are very broad in medium due to phonon excitation of the noble gas lattice, and are on the order of $10^3-10^4$ GHz or roughly $1-10$ nm, which is orders of magnitude larger than in vacuum.

As an example, the most abundant potential background source will be components of air trapped as impurities in the noble gas film. Noble gases are commercially available with ppm purities, and can be further purified to ppb levels with gettering. For an Argon film of 100 $\mu$m thickness, there will be roughly $2\times10^{20}$ Argon atoms/cm$^{2}$, with $10^{14}$ molecules of air assuming ppm purity. To achieve $S/B=1$ would require an off-resonant suppression factor of roughly $10^{-28}$, a distance of almost 5 linewidths from resonance for a Gaussian absorption lineshape. As the molecular components of air do not absorb until well into the ultraviolet, they are not expected to contribute significantly to the background rate, with the exception of O$_2$ near 763 nm.


\section{Limitations and Summary}

Several limitations exist for the SAM detection scheme. First, while atomic species can be selectively excited, this method is incapable of distinguishing between different isotopes of the same species. Although small isotope differences exist in atomic hyperfine structure due to the nuclear spin, the linewidths in medium are so broad that any isotopic variation becomes obscured. Second, while this method is potentially applicable to a wide range of species, species without optically accessible transitions cannot be detected, which eliminates the noble gas elements along with elements like Fluorine, whose lowest lying transition occurs at 97.7 nm in vacuum. Third, it is not suited to detect products that are abundant in a vacuum system, such as Oxygen, Nitrogen, or Carbon, as it would be impossible to grow a solid noble gas film without thousands of such contaminant atoms even with the highest achievable noble gas purity. Fourth, the detection mechanism is slow for the SAM relative to traditional methods since the products are not detected immediately after creation, but rather at a later time when they are imaged. Therefore short lived isotopes ($\tau_{1/2}<1$ day) are not suitable unless the daughter nuclei are also optically detectable and the daughters are a different species from the beam atoms and any background atoms.

To summarize, important reactions for nucleosynthesis processes are often difficult to measure because of their low yield, whether it be due to extremely low cross sections, low intensity rare isotope beams, or high background rates. We propose a novel detection method for low yield nuclear reactions that captures the product atoms in a cryogenically frozen film of a solid noble gas where they are optically imaged and counted. This method can offer near unity capture and detection efficiency, feasibly achieve single atom sensitivity, and is potentially applicable to many astrophysically important nuclear reactions. The chief advantage of the Single Atom Microscope is that it is not sensitive to neutron, gamma, or charged particle background sources, and could therefore outperform traditional detection methods.


\section{Acknowledgements}

Special thanks to Sam Austin, Artemis Spyrou, Hendrik Schatz, and Xiao-Dong Tang for helpful duscussions. This work benefited from support by the National Science Foundation under Grant No. PHY-1430152 (JINA Center for the Evolution of the Elements). This work is supported by Michigan State University, the Director’s Research Scholars Program at the National Superconducting Cyclotron Laboratory, and the U.S. National Science Foundation under grant number \#1654610.


\bibliography{SAM-refs-prc}

\end{document}